\documentclass[twocolumn,secnumarabic,amssymb, nobibnotes, aps, prd, superscriptaddress]{revtex4-1}
\usepackage{graphicx}
\usepackage{braket}
\usepackage{array}
\usepackage{amsmath}
\usepackage{here}

\begin{document}

\title{Phonon-number-resolving Detection of Multiple Local Phonon Modes in Trapped Ions}%

\author{Ryutaro Ohira}%
\email{u696585a@ecs.osaka-u.ac.jp}
\affiliation{Graduate School of Engineering Science, Osaka University, 1-3 Machikaneyama, Toyonaka, Osaka, Japan}
\author{Takashi Mukaiyama}%
\affiliation{Graduate School of Engineering Science, Osaka University, 1-3 Machikaneyama, Toyonaka, Osaka, Japan}
\affiliation{Quantum Information and Quantum Biology Division, Institute for Open and Transdisciplinary Research Initiatives, Osaka University, 1-3 Machikaneyama, Toyonaka, Osaka, Japan}
\author{Kenji Toyoda}%
\affiliation{Quantum Information and Quantum Biology Division, Institute for Open and Transdisciplinary Research Initiatives, Osaka University, 1-3 Machikaneyama, Toyonaka, Osaka, Japan}

\date{\today}

\begin{abstract}
We propose and demonstrate phonon-number-resolving detection of the multiple local phonon modes in a trapped-ion chain. To mitigate the effect of phonon hopping during the detection process, the probability amplitude of each local phonon mode is mapped to the auxiliary long-lived motional ground states. Sequential state-dependent fluorescence detection is then performed. In the experiment, we have successfully observed the time evolution of two local phonon modes in two ions, including the phonon-number correlation between the two modes.
\end{abstract}


\maketitle


Trapped ions provide a well-controlled physical system that is applicable to the processing of quantum information and the investigation of many-body physics. Phonons in trapped ions have controllability via optical manipulation in a similar way to that of internal states. In addition, phonons have large information capacity, due to the high Hilbert-space dimensions of the phonon modes. Phonons can be used for the simulation of physical systems involving Bose particles, such as the Bose-Hubbard \cite{porras2004bose}, spin-boson \cite{porras2008mesoscopic,lemmer2018trapped} or Holstein models \cite{mezzacapo2012digital}, and molecular vibrations \cite{huh2015boson,shen2018quantum}.

For these applications, the concept of local phonons is important. When multiple ions are confined in a potential and the confinement force in one direction acting on each ion is much stronger than the Coulomb forces, the vibrational mode of the ion is almost independent of the vibrational modes of other ions, and still the modes are weakly coupled via the Coulomb interactions. Phonons in these modes are termed as \textit{local phonons}; they possess particle-like characteristics, exhibit behaviour such as hopping \cite{harlander2011trapped,brown2011coupled,haze2012observation,toyoda2015hong,debnath2018observation,tamura2019quantum}, and their total numbers are conserved. These characteristics are advantageous for many applications including quantum simulation.

The hopping of a local phonon, where a single quantum of vibration hops between ions mediated by Coulomb interactions, has been observed to date \cite{harlander2011trapped,brown2011coupled,haze2012observation,toyoda2015hong,debnath2018observation,tamura2019quantum}. Applications for quantum information processing and quantum simulation using local phonon modes have been proposed \cite{zhu2006arbitrary,deng2008quantum,ivanov2009simulation} and some of them have been realized \cite{toyoda2013experimental}. The dynamics of the energy transport have also been investigated using the propagation of locally excited phonons in trapped ions \cite{ramm2014energy,abdelrahman2017local}.

The hopping of multiple phonons in an ion chain could offer a platform to realize advanced quantum simulation and information processing which cannot be addressed with a single local phonon. One of the applications of multi-phonon hopping is boson sampling \cite{aaronson2011computational,broome2013photonic,spring2013boson,tillmann2013experimental,crespi2013integrated,wang2017high,he2017time,loredo2017boson}, which has been of growing interest in terms of the demonstration of quantum supremacy. Characteristics such as deterministic preparation, efficient read-out and universal phonon-mode mixing mean that a system of phonons in trapped ions is expected to be a promising platform for the scalable physical implementation of boson sampling \cite{shen2014scalable}. Multiple bosonic quanta and their detection are also essential requirements for the simulation of molecular transitions and dynamics \cite{huh2015boson,shen2018quantum,sparrow2018simulating}.

One of the key elements for the efficient realization of such applications is particle-number-resolving detection over multiple modes. For example, in the case of boson sampling, when single-quantum detectors (such as single-photon detectors) are used for detection, the preparation of bosonic particles with sufficiently low density over many modes is required to avoid the coincidence of multiple particles at each output port (boson birthday bound \cite{aaronson2011computational}), which necessitates a large overhead with respect to the number of modes for a given number of bosonic particles. If particle-number-resolving detectors are employed instead of single-quantum detectors, this restriction can be largely relaxed. As far as we are aware, the Fock-state-based projective measurement of multi-phonon hopping over multiple modes has not been reported to date. The hopping of two phonons in a two-ion chain that involves Hong-Ou-Mandel interference has been reported recently \cite{toyoda2015hong}. However, the probability amplitude of the local phonon modes has been inferred from the phonon coincidence, rather than by a phonon-number-resolving approach \cite{toyoda2015hong}.

\begin{figure*}[t]
\centering
  \includegraphics[width=16.5cm]{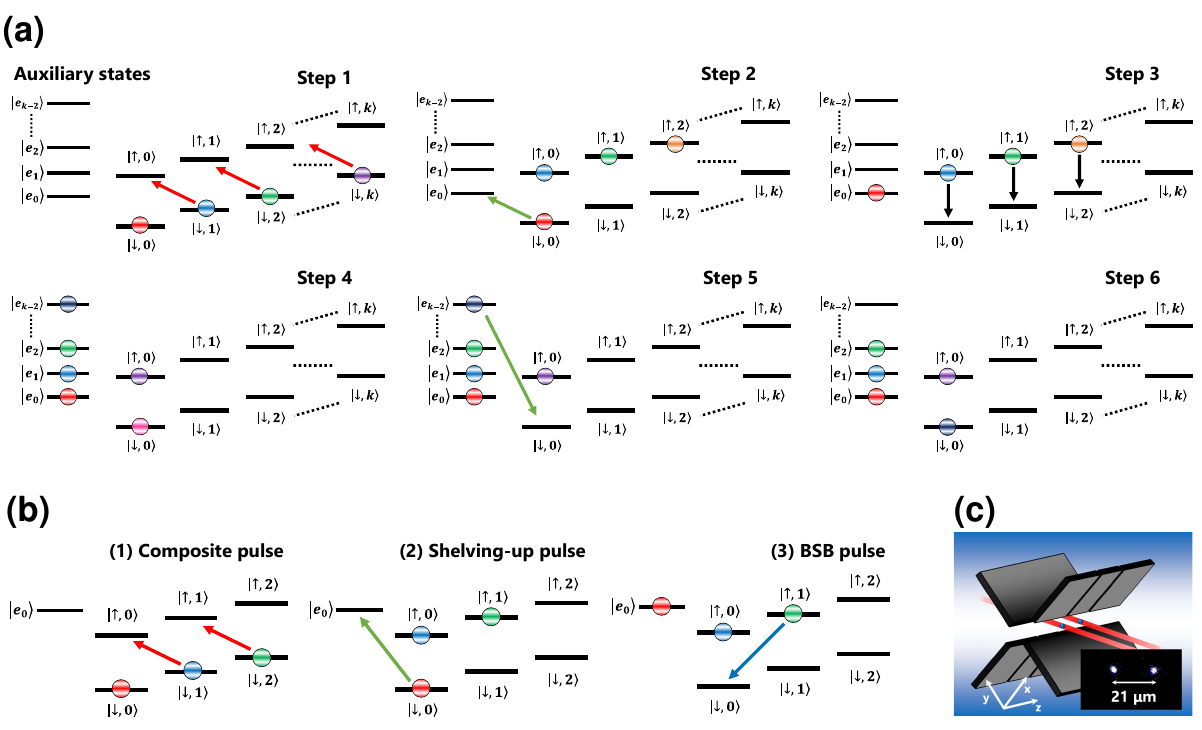}
\caption{\label{fig1} (a) Phonon-number-resolving detection of multiple local phonon modes. The sequence for the $\it{j}$-th ion in an $\it{N}$-ion chain is shown schematically. In Step 1, adiabatic passage over the red sideband transition is applied to the ions. The probability amplitude of each local phonon mode is transferred from $\ket{\downarrow,n_{j}}$ to $\ket{\uparrow,n_{j}-1}$ ($n_{j}$$\geq$1). In Step 2, the probability amplitude in $\ket{\downarrow,n_{j}=0}$ is transferred to one of the long-lived auxiliary states. In Step 3, $\pi$ pulses at the carrier transition are applied. Steps 1, 2 and 3 are then repeated for $\it{k}$-1 times. After repeating these steps, every probability amplitude of the local phonon mode is mapped to the motional ground states. Detection of the state-dependent fluorescence is then conducted (Step 4). If no fluorescence is detected, then the probability amplitude of $\ket{k-2}$ in $\ket{e_{k-2}}$ is transferred to $\ket{\downarrow,n_{j}=0}$ and fluorescence detection is performed (Step 5). Steps 4 and 5 are repeated until fluorescence is detected (Step 6). If fluorescence is detected, then the quantum state is determined and the entire sequence for the $\it{j}$-th ion is stopped. (b) Simplified mapping process used in the experiment. The composite pulse technique has been used instead of adiabatic passage. The probability amplitude of $\ket{n_{y}=0}$ is transferred to $\ket{D_{5/2},m_{j}=-5/2}\equiv\ket{e_{0}}$ (shelving-up pulse). A blue-sideband (BSB) $\pi$ pulse is then applied. (c) Schematic illustration of the experimental setup. Two $^{40}{\rm Ca}^{+}$ ions are trapped in a linear Paul trap. The distance between the ions is approximately 21 $\rm{\mu}m$. Two tightly focused beams are used to manipulate the local phonons.}
\end{figure*}

Phonon-number-resolving detection for a single ion has already been realized \cite{an2015experimental,um2016phonon,wolf2019motional}. The detection schemes are based on sequential adiabatic passage over sideband transitions and spin- or state-dependent fluorescence detection. However, the local phonons are not the eigenstates of the total Hamiltonian for the system, and hence phonon hopping can occur during each measurement process; therefore, it is not possible to simply extend these phonon-number-resolving detection schemes to the case of ion chains with multiple local phonon modes. (The high-fidelity measurement takes milliseconds, while the hopping period is typically around a few hundred microseconds considering realistic experimental parameters.)

In this letter, we propose and demonstrate phonon-number-resolving detection of the multiple local phonon modes in an ion chain. To prevent the influence of phonon hopping on the measurement results, the probability amplitude of each local phonon mode is rapidly mapped to the auxiliary internal states. After the mapping process, sequential state-dependent fluorescence detection is performed, which allows for the detection of correlations among the Fock states of different modes. As a demonstration, we have performed the Hong-Ou-Mandel experiment using local phonon modes \cite{toyoda2015hong,hong1987measurement}. The time evolution of the two-mode local phonons was seccessfully observed using the proposed phonon-number-resolving detection.


We first describe a general scheme for phonon-number-resolving detection [Fig.\,1(a)]. Let us consider $\it{N}$ ions that are trapped in a harmonic potential forming a linear chain. The quantum state of the $\it{j}$-th ion is assumed to be $\ket{\psi_{j}}=\sum_{n_{j}=0}^{k}c_{n_{j}}\ket{n_{j}}$, where $c_{n_{j}}$ is the probability amplitude that satisfies $\sum_{n=0}^{k}|c_{n_{j}}|^2=1$. $\ket{n_{j}}$ is the Fock state of the radial local phonon modes of the $\it{j}$-th ion. In Step 1, adiabatic passage over the red sideband transition \cite{watanabe2011sideband,shen2014scalable,an2015experimental,um2016phonon,gebert2016detection,wolf2019motional} is applied to the ions. The probability amplitude of each local phonon mode is transferred from $\ket{\downarrow,n_{j}}$ to $\ket{\uparrow,n_{j}-1}$ ($n_{j}$$\geq$1). In Step 2, the probability amplitude in $\ket{\downarrow,n_{j}=0}$ is transferred to one of the long-lived auxiliary states. In Step 3, $\pi$ pulses at the carrier transition are applied. Steps 1, 2 and 3 are repeated $\it{k}$-1 times. Every probability amplitude of the local phonon mode is then mapped to the motional ground states. The detection of the state-dependent fluorescence is then conducted (Step 4). If no fluorescence is detected, then the probability amplitude of $\ket{k-2}$ in $\ket{e_{k-2}}$ is transferred to $\ket{\downarrow,n_{j}=0}$ and fluorescence detection is performed (Step 5). Steps 4 and 5 are repeated until the fluorescence is detected (Step6). If fluorescence is detected, then, the quantum state is determined and the entire sequence for the $\it{j}$-th ion is stopped.


The scheme described here can be applied to general cases with large phonon numbers under ideal conditions (sufficiently fast transfer with respect to the hopping time). On the other hand, under realistic conditions with finite operation time, the number of operations and thus the maximum phonon number may be bounded in relation to the desired fidelity of operation (see discussion given later for further details).

If the maximum phonon number per mode is restricted to two, then a (non-general) simpler scheme can be used, which enables faster operations with less sacrifice of fidelity. This scheme is sufficient for the detection of phonon-number correlations in Hong-Ou-Mandel interference\cite{toyoda2015hong, shen2018quantum}, and allow extensions to more modes without an increase in the number of operations. Below we present experimental procedures based on such a simplified scheme and the results obtained.

The mapping procedure for the simplified scheme of phonon-number-resolving detection is shown in Fig.\,1(b). We have used the composite pulse technique \cite{levitt1979nmr,gulde2003implementation,schmidt2003realization} instead of adiabatic passage. The Rabi frequency at the red-sideband transition between $\ket{\uparrow,n_{y}-1}\leftrightarrow\ket{\downarrow,n_{y}}$ is $\Omega_{\rm RSB}=\Omega_{0}\sqrt{n_{y}}\eta$, where $\eta$ is the Lamb-Dicke parameter and $\Omega_{0}$ is the Rabi frequency at the carrier transition. The composite-pulse sequence at the red-sideband transition was implemented to compensate for the Rabi frequency difference between $\ket{\uparrow,n_{y}=0}\leftrightarrow\ket{\downarrow,n_{y}=1}$ and $\ket{\uparrow,n_{y}=1}\leftrightarrow\ket{\downarrow,n_{y}=2}$. The operation for the red-sideband transition for the {\it j}-th ion can be expressed as:
\begin{equation}
  R_{\rm RSB}(\theta, \phi) = \exp\Biggl[i\frac{\it{\theta}}{\rm 2}(e^{i\phi}\sigma^{+}\hat{a}_{j,y}^{\dagger}+e^{-i\phi}\sigma^{-}\hat{a}_{j,y})\Biggr], 
\end{equation}
where $\theta$ and $\phi$ denote the angle of the qubit rotaion and the rotation axis, respectively. The qubit rotation angle $\theta$ is ${\Omega_{\rm RSB}}t$, where $t$ is the pulse duration. The red-sideband composite-pulse sequence consists of three consecutive pulses:
\begin{equation}
  R_{\rm composite} = R_{\rm RSB}\biggl(\frac{\pi}{2}, 0\biggr)R_{\rm RSB}\biggl(\frac{\pi}{\sqrt{2}}, \frac{\pi}{2}\biggr)R_{\rm RSB}\biggl(\frac{\pi}{2}, 0\biggr).
\end{equation}
The probability amplitude in $\ket{\downarrow,n_{y}=0}$ is transferred to a long-lived auxiliary state $\ket{e_{0}}$, and a blue-sideband $\pi$ pulse between $\ket{\uparrow,n_{y}=1}$ and $\ket{\downarrow,n_{y}=0}$ is applied. The read-out procedure is performed in a similar way to Steps 4, 5 and 6.

Here, we have implemented phonon-number-resolving detection for a single ion and two ions. The experimental setup is schematically shown in Fig.\,1(c), where two $^{40}{\rm Ca}^{+}$ ions are trapped in a linear Paul trap. The distance between the ions is approximately 21 $\rm{\mu}m$, where the axial secular frequency is $ \omega_{z}\approx $ 2$\pi\times$150 kHz. The radial secular frequencies are $(\omega_{x}, \omega_{y})=$ 2$\pi\times$(3.17, 3.00) MHz. Internal states $\ket{S_{1/2},m_{j}=-1/2}\equiv\ket{\downarrow}$ and $\ket{D_{5/2},m_{j}=-1/2}\equiv\ket{\uparrow}$ are used to encode the spin states. In addition,$\ket{D_{5/2},m_{j}=-1/2}\equiv\ket{e_{0}}$ is used as the auxiliary state. The Zeeman sublevels of the $D_{5/2}$ state have a lifetime of $\sim$ 1 s.

The experiment starts with Doppler cooling and motional ground state cooling. The ion is Doppler cooled using the transitions ${\it S_{\rm 1/2}}$-${\it P_{\rm 1/2}}$ (397 nm) and ${\it D_{\rm 3/2}}$-${\it P_{\rm 1/2}}$ (866 nm). The two radial directions, {\it x} and {\it y}, are then cooled to the motional ground state by resolved sideband cooling. The narrow quadrupole transition, ${\it S_{\rm 1/2}}$-${\it D_{\rm 5/2}}$, is used for the ground state cooling. In the experiment, the local phonon mode along {\it y} direction is manipulated. 

The experiment was performed with a single ion to confirm the validity of the proposed phonon-number-resolving detection. The ion was prepared in either of three different Fock states, $\ket{n_{y}=0}$, $\ket{n_{y}=1}$ or $\ket{n_{y}=2}$, using the repetition of a blue-sideband $\pi$ pulse followed by a $\pi$ pulse at the carrier transition. A 854 nm laser was also applied after each carrier $\pi$ pulse to clear out the probability amplitude in the excited state. The Fock state is subsequently measured by phonon-number-resolving detection. The total time required for the composite-pulse sequence under the current conditions is 40 $\mu$s, which is much shorter than half of the hopping period ($\pi/\kappa\approx170$ $\mu$s, where $\kappa\approx2\pi\times3$ kHz), so that the effect of hopping during this time is not significant. All the other pulses that are applied on the carrier transtions are approximately 1 $\mu$s; therefore, the effect of hopping during those pulses can be neglected.


The experimental results for a single ion are shown in Fig.\,2. The result when the initial state is prepared in $\ket{n_{y}=0}$, $\ket{n_{y}=1}$ or $\ket{n_{y}=2}$ corresponds to (a), (b) and (c) of Fig.\,2, respectively. The red bars represent the experimental results and each result is the average of 500 experiments. The black bars are the probability amplitude for each local phonon obtained by fitting the blue-sideband Rabi oscillation. The results, clearly indicate that the proposed phonon-number-resolving detection is working for a single ion.

\begin{figure}[t]
\centering
  \includegraphics[width=8.5cm]{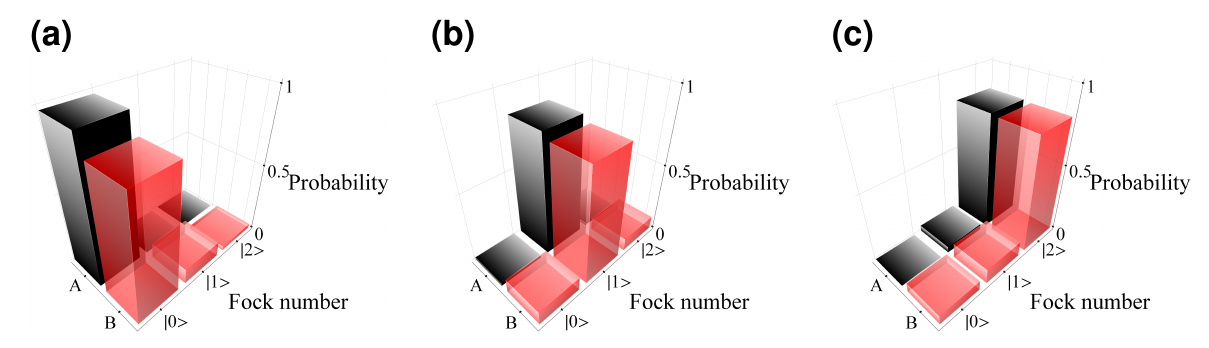}
\caption{\label{fig2} Phonon-number-resolving detection of a single ion. Results for initial states of (a) $\ket{n_{y}=0}$, (b) $\ket{n_{y}=1}$, and (c) $\ket{n_{y}=2}$. Labels A and B represent data obtained from fitting of the blue-sideband Rabi oscillation (black bar) and phonon-number-resolving detection (red bar), respectively.}
\end{figure}

\begin{figure}[t]
\centering
  \includegraphics[width=8.5cm]{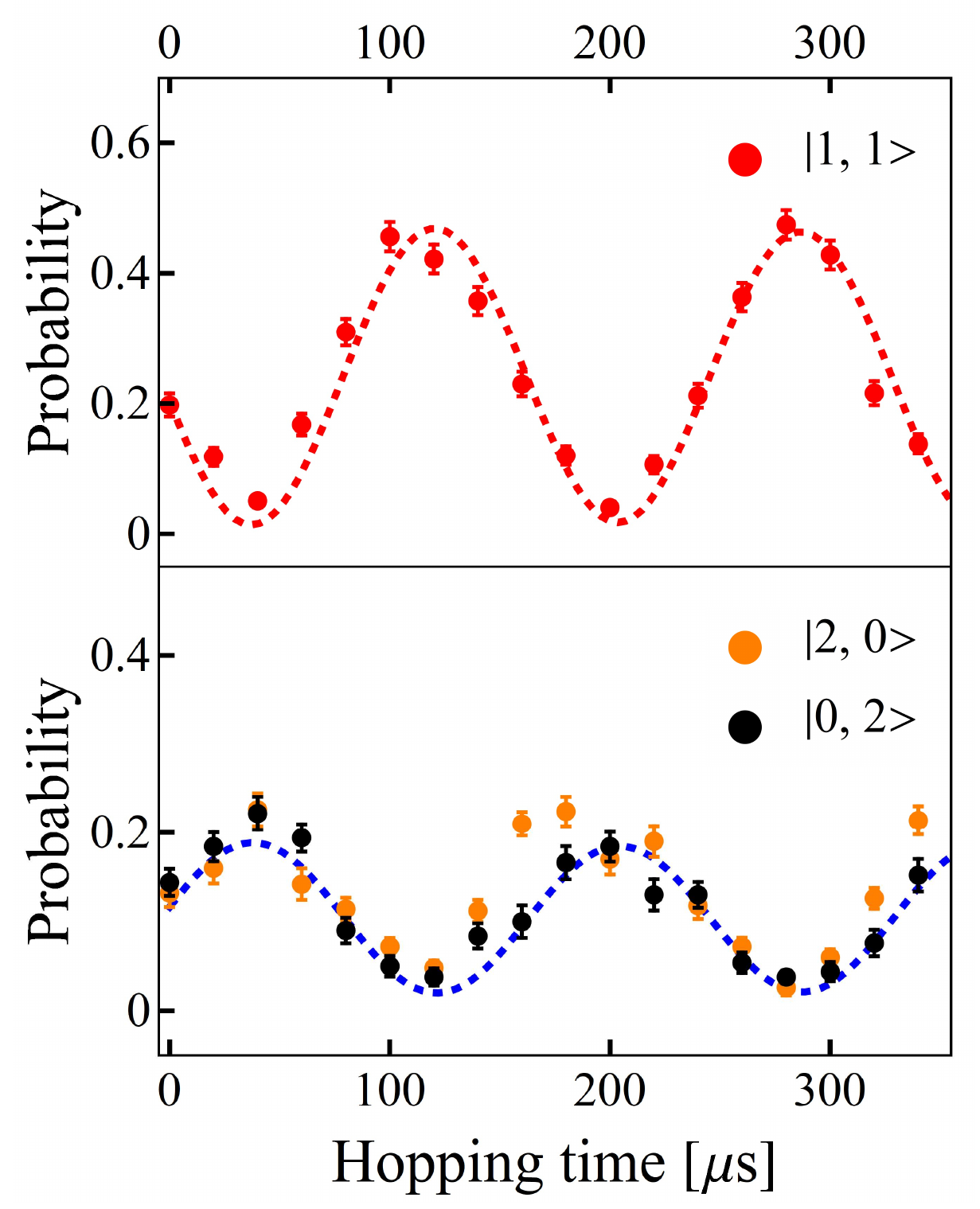}
\caption{\label{fig3}Results of phonon-number-resolving detection for two ions with two phonons. Red, orange and black circles represent the measured probability amplitude of $\ket{1, 1}$, $\ket{2, 0}$ and $\ket{0, 2}$, where the first and second numbers represent the phonon numbers for the first and second ions, respectively. Each time step corresponds to the collection of 500 experiments. The error bars represent the statistical uncertainties of 1$\sigma$. The dashed curves represent the numerical calculation of each local phonon mode. The simulation was performed based on the Hamiltonian given in Eq.\,(3).}
\end{figure}

Phonon-number-resolving detection was implemented for two ions with two phonons. The experimental sequence is almost the same as that for the single ion experiment. The initial state preparation was comprised of Doppler cooling, ground state cooling and manipulation of the local phonon mode using the sideband transition. As an initial state, the ions were prepared in the state $\ket{\psi_{\rm init}}=\ket{\downarrow,n_{y}=1}\otimes\ket{\downarrow,n_{y}=1}$. After initial state preparation, the phonon hopping is observed.

The Hamiltonian of the system can be represented as \cite{porras2004bose,deng2008quantum,ivanov2009simulation}.
\begin{equation}
  H = \sum_{j=\rm 1,2}(\omega_{y}-\frac{\it{\kappa}}{\rm 2})\hat{a}_{j,y}^{\dagger}\hat{a}_{j,y}+\frac{\it{\kappa}}{\rm 2}(\hat{a}_{1,y}\hat{a}_{2,y}^{\dagger}+\hat{a}_{1,y}^{\dagger}\hat{a}_{2,y}), 
\end{equation}
where $\hbar=1$ is used, $\it{\kappa}=e^{\rm 2}/({\rm 4}\pi\varepsilon_{0}\it{m}\it{d}^{\rm 3}\omega_{y})$ is the hopping rate, {\it m} and {\it e} are the mass and charge of the ion, respectively, and $\it{d}$ represents the distance between the ions. Using the parameters in this experiment, the hopping rate can be calculated as $\it{\kappa}\approx$ 2$\pi\times3\,{\rm kHz}$. $\hat{a}_{j,y}$ and $\hat{a}_{j,y}^{\dagger}$ are the annihilation and creation operators of the local phonon mode along the {\it y} direction. Phonon-number-resolving detection is then performed after the hopping period.


The experimental results are given in Fig.\,3. Each of the local phonon mode probability amplitudes is denoted as $\ket{1, 1}$ (red circles), $\ket{2, 0}$ (orange circles) and $\ket{0, 2}$ (black circles), where the first and second numbers represent the phonon numbers for the first and second ions, respectively. Each time step corresponds to the collection of 500 experiments. The dashed curves represent the numerically calculated probability amplitude of each local phonon mode. The simulation is performed based on the Hamiltonian in Eq.(3), including the sideband Rabi oscillation decay rate. This experiment corresponds to a Hong-Ou-Mandel experiment using the local phonon modes of trapped ions \cite{toyoda2015hong}. The quantum bosonic interference is clearly observed in Fig.\,3. The contrast of the measurement results are limited by the lack of coherence of the sideband Rabi oscillation. Therefore, the fidelity of the phonon-number-resolving detection is expected to be improved simply by a reduction of the decoherence of the sideband Rabi oscillation.


In principle, the phonon-number-resolving detection presented here is a scalable scheme in terms of the number of ions. The scalability in terms of the phonon number is determined by the number of available internal states and that of sideband transfers that can be performed before the effect of hopping become prominent. In the case of $^{40}{\rm Ca}^{+}$ ions, Zeeman sublevels of the ${D_{5/2}}$ state are suitable for the auxiliary states for mapping. All the Zeeman sublevels in ${S_{1/2}}$ and ${D_{5/2}}$ are accessible using the optical transition and rf transitions. Therefore, phonon-number-resolving detection for up to eight phonon numbers can be realized. If ions with non-zero nuclear spin are used, such as $^{43}{\rm Ca}^{+}$ ion with nuclear spin $7/2$, then, it is possible to use large numbers of internal states to store probability amplitudes. 

The generalized scheme for the implementation of phonon-number-resolving detection is enabled with the use of adiabatic passage. Using adiabatic passage over sideband transitions can realize the highly efficient and uniform transfer of local-phonon probability amplitudes over a wide range of phonon numbers \cite{watanabe2011sideband,shen2014scalable,an2015experimental,um2016phonon,gebert2016detection,wolf2019motional}. However, due to the relatively long pulse duration of adiabatic passage, it is not possible to simply extend the scheme to multi-phonon hopping. To precisely observe the quantum dynamics of the local phonon modes, the probability amplitude of the local phonon modes must be rapidly mapped to the auxiliary states.

Recently, transitionless quantum driving has been proposed to speed up adiabatic operations \cite{demirplak2003adiabatic,berry2009transitionless,chen2010shortcut}. Transitionless adiabatic passage has been realized in a trapped-ion system \cite{um2016phonon,an2016shortcuts}. By combining this advanced technique, the problem of long operation time could be mitigated. We have performed a numerical simulation of the probability transfer using transitionless quantum driving and confirmed that adiabatic passage over the red-sideband transition that takes 70 $\rm{\mu}$s can transfer eight phonon number states including $\ket{n_{y}=0}$, $\ket{n_{y}=1}$, $\ket{n_{y}=2}$,$\cdot$$\cdot$$\cdot$, and $\ket{n_{y}=7}$ with more than 99\% fidelity. In the numerical simulation, 40 kHz was used for the Rabi frequency at the red-sideband transition. Note that the decoherence of the sideband Rabi oscillation and the effect of the AC stark shift were neglected. 

In order to consider the experimental feasibility of phonon-number-resolving detection for multiple phonon modes, we suppose the following situation. Two $^{40}{\rm Ca}^{+}$ ions are trapped and the distance between them is 40 $\rm{\mu}$m. The secular frequency along the radial direction is 3.0 MHz. The phonon hopping rate is then calculated to be approximately 0.5 kHz, and half of the hopping period is 1 ms. Several repetitions of adiabatic passage that take 70 $\rm{\mu}$s can be performed within this time. Therefore, implementation of the generalized scheme may be realistic in such a case.

In conclusion, we have proposed and demonstrated the phonon-number-resolving detection of multiple local phonon modes. This phonon-number-resolving detection was implemented for a two-ion chain with two phonons and the time evolution of the multiple local phonon modes was successfully observed.

This work was supported by MEXT Quantum Leap Flagship Program (MEXT Q-LEAP) Grant Number JPMXS0118067477.

\bibliography{ref}
\bibliographystyle{apsrev4-2}

\end{document}